\documentclass[12pt]{article}
\usepackage{amssymb}
\usepackage{epsfig}
\begin{document}
%
\setlength{\baselineskip}{0.65cm}
\setlength{\parskip}{0.35cm}
\renewcommand{\thesection}{\Roman{section}}
%
\begin{titlepage}
%
\begin{flushright}
BNL-NT-04/24 \\
RBRC-423 \\
July 2004
\end{flushright}

\vspace*{1.1cm}
\begin{center}
\LARGE

{\bf {Observations on dA scattering}}\\[2mm]

\medskip
{\bf {at forward rapidities}}\\

\vspace*{2.5cm}
\large 
{V.\ Guzey$^a$, M.\ Strikman$^b$, and W.\ Vogelsang$^c$}

\vspace*{0.5cm}
\normalsize
{\em $^a$ Institut f\"{u}r Theoretische Physik II, 
Ruhr-Universit\"{a}t Bochum, D-44780 Bochum, Germany}\\

\vspace*{0.5cm}
\normalsize
{\em $^b$Department of Physics, Pennsylvania State University,
University Park, PA, U.S.A.}\\

\vspace*{0.5cm}
{\em $^c$Physics Department and RIKEN-BNL Research Center, \\
Brookhaven National Laboratory,
Upton, New York 11973, U.S.A.}\\

\end{center}

\vspace*{2.0cm}
\begin{abstract}
\noindent
We point out that the suppression in the ratio $R_{{\mathrm{dAu}}}$
recently observed by the {\sc Brahms} collaboration in forward
scattering is stronger than usually appreciated. This is related 
to the fact that at forward rapidities {\sc Brahms} measures
negatively charged hadrons and that $R_{{\mathrm{dAu}}}$ is
defined from the ratio of dA and $pp$ scattering cross sections.
We also investigate the influence of standard shadowing on
$R_{{\mathrm{dAu}}}$, and the typical values of partonic momentum 
fractions relevant in forward scattering. We find that $x_{\mathrm{Au}}
\ge 0.02 $ dominate in the  cross section.
\end{abstract}
\end{titlepage}
\newpage
%
\section{Introduction}
%
The {\sc Brahms} collaboration has recently presented
measurements of the ratio $R_{{\mathrm{dAu}}}$
of hadron production cross 
sections in dAu and $pp$ collisions~\cite{brahms}. With increasing 
hadron rapidity, the data indicate a growing suppression
of the ratio. Theoretical analyses of the data have
focused on nuclear effects related to the gold nucleus, 
emphasizing variously the role of small $x$ physics in the nuclear 
production~\cite{cgcmodels}, as well as of nuclear-enhanced power 
corrections in the dA cross section \cite{qv}. Other
studies~\cite{ds,rvogt} have addressed the question of whether 
leading-twist shadowing could be responsible for the 
observed suppression.

In the present paper we hope to add valuable information
that will have an impact on the present and future attempts 
to interpret the {\sc Brahms} data, and on plans for further measurements.
We base our analysis on next-to-leading order (NLO) leading-twist 
perturbative-QCD (pQCD) calculations
of inclusive hadron production. Such calculations have 
enjoyed considerable success in comparisons with data
from $pp$ collisions at RHIC at $\sqrt{s}=200$~GeV. 
They yield good agreement with the {\sc Phenix} 
data~\cite{phenix} for $pp\to \pi^0 X$ at central 
rapidities, and with data from {\sc Star}~\cite{star} on 
$pp\to \pi^0 X$ at forward rapidities $\eta=3.8$ and $\eta=3.3$, 
the latter being essentially in the kinematic range explored by
{\sc Brahms} in their most forward measurements. It is appropriate
to point out that there are sizable uncertainties in the NLO
calculation, related to the choice of fragmentation functions
and scales. However, we are confident that for the kinematics 
relevant for {\sc Brahms} NLO pQCD does explain at least 
50\% of the cross section, as a conservative estimate,
and hence is a viable tool for obtaining deeper insights into the
production mechanism, at least for $pp$ collisions.  

The main point of our analysis can be stated very simply:
the nuclear effect reported by {\sc Brahms} actually appears to be 
substantially larger than appreciated in the 
studies~\cite{cgcmodels,qv,rvogt}.
This is related to the fact that in the very forward region,
at rapidities $\eta=2.2$ and $\eta=3.2$, {\sc Brahms}
only measured {\it negatively charged} hadrons ($h^-$) and
not the charge average $(h^+ + h^-)/2$ as at the more central 
rapidities. In the ratio $R_{{\mathrm{dA}}}$ the denominator
refers to $pp$ collisions, and negatively charged hadrons
are expected to be produced more rarely in $pp$ than in dA
collisions, for 
which from isospin considerations it follows that at least $\pi^+$ 
and $\pi^-$ should be produced in equal numbers. This 
immediately implies that the ratio of ${\mathrm{dAu}}\to h^- X$
and $pp\to h^- X$ cross sections is ``intrinsically enhanced'',
by actually a factor of about 1.5, as we will show.
The fact that {\sc Brahms} nonetheless reports a suppression
of the ratio is therefore truly remarkable and awaits 
further investigation.

We also revisit in our analysis the following questions:
\begin{description}
\item[(i)] what are the most relevant parton momentum fractions $x$
for hadron production at {\sc Brahms}, in particular at very 
forward rapidity $\eta=3.2$ where the suppression of $R_{{\mathrm{dA}}}$
is largest? To what extent are truly small $x$, say, $x<10^{-3}$ probed?
\item[(ii)] how relevant is leading-twist nuclear shadowing for the 
explanation of the {\sc Brahms} data?
\end{description}
These questions have already been addressed in some detail 
in~\cite{rvogt}. Our analysis extends that study by
providing results within a full NLO calculation. This will
generally lead to more reliable results. In addition, the 
enhancement effect mentioned above has of course also direct 
implications for estimates for $R_{{\mathrm{dA}}}$ obtained when
using leading-twist nuclear shadowing. Our calculations 
therefore provide an improved estimate as compared to the
results of~\cite{rvogt}, where the enhancement was not taken
into account. We furthermore explore more thoroughly the possible effects of
leading-twist nuclear shadowing by making more extreme assumptions 
on the structure of screening at intermediate $x$, tolerated because
of possible uncertainties in the connection between diffractive HERA data 
and gluon shadowing.

In sec.~\ref{elementary} we discuss the ranges of partonic
momentum fractions mainly probed by the forward {\sc Brahms} data.
Section \ref{nuclshad} addresses the leading-twist nuclear 
shadowing and its relevance in forward dA scattering. 
With the findings of sections~\ref{elementary} and
\ref{nuclshad}, we are in the position to discuss 
$R_{{\mathrm{dA}}}$ in the forward region in more detail.
This is done in Section~\ref{isospin}, where we 
emphasize our main point related to the normalization of 
$R_{{\mathrm{dA}}}$ by the $pp\to h^- X$ cross section. 
We summarize and conclude with section \ref{conclusions}.  

%
\section{Kinematics and $x$ ranges probed in forward scattering}
%
\label{elementary}
We consider the reaction $H_1 H_2\to h X$, where $H_1, H_2$ 
are initial hadrons and $h$ is a hadron in the final state produced at 
high transverse momentum $p_T$. Since large $p_T$ ensures large 
momentum transfer, the cross section for the process may be written 
in a factorized form,
\begin{eqnarray}
\label{eq1}
d\sigma &=&\sum_{a,b,c}\, 
\int_{x_2^{\mathrm min}}^1 dx_2
\int_{x_1^{\mathrm min}}^1 dx_1 
\int_{z^{\mathrm min}}^1 dz \,\,
f_a^{H_1} (x_1,\mu) \, f_b^{H_2} (x_2,\mu) \,
D_c^h(z,\mu) \, \nonumber \\ [2mm]
&&
\times \, 
d \hat{\sigma}_{ab}^{c}(x_1 P_{H_1}, x_2 P_{H_2}, P_h/z, \mu) \; ,
\end{eqnarray}
where the sum is over all contributing partonic channels $a+b\to
c + \ldots$, with $d\hat{\sigma}_{ab}^{c}$ the associated
short-distance cross section which may be evaluated
in QCD perturbation theory:
\begin{equation}
d\hat{\sigma}_{ab}^{c}\,=\,
d\hat{\sigma}_{ab}^{c,(0)}+
\frac{\alpha_s}{\pi} d\hat{\sigma}_{ab}^{c,(1)}+\ldots \; .
\end{equation}
The leading-order (LO) contributions $d\hat{\sigma}_{ab}^{c,(0)}$ 
are of order $\alpha_s^2$; the next-to-leading order (NLO) corrections
are known~\cite{pionnlo} and will be included in our analysis. 

In Eq.~(\ref{eq1}), $f_i^H (x,\mu)$ denotes the distribution 
function at scale $\mu$ for a parton of type $i$ in hadron $H$, 
carrying the fraction $x$ of the hadron's light-cone momentum. 
Likewise, $D_c^h(z,\mu)$ describes the fragmentation of produced 
parton $c$ into the observed hadron $h$, the latter taking momentum 
fraction $z$ of the parton momentum. The scale $\mu$ in Eq.~(\ref{eq1})
stands generically for the involved renormalization and factorization 
scales. $\mu$ should be of the order of the hard scale in
the process; in the following we choose $\mu=p_T$. The dependence 
on $\mu$ is actually quite large even at NLO~\cite{pionnlo}; however, in this
work we are mainly interested in ratios of cross sections
for which the $\mu$ dependence is fairly insignificant. 

The lower limits of the integrations over momentum fractions
in Eq.~(\ref{eq1}) may be derived in terms of $x_T=2p_T/\sqrt{s}$ 
and the pseudorapidity $\eta$ of the produced hadron. They are 
given by
\begin{eqnarray} \label{eq2}
x_2^{\mathrm min} &=& \frac{x_T \, {\mathrm e}^{-\eta}}{2-x_T \, 
{\mathrm e}^{\eta}}
\; , \nonumber \\
x_1^{\mathrm min} &=& \frac{x_2\, x_T \, {\mathrm e}^{\eta}}{2 x_2-x_T \, 
{\mathrm e}^{-\eta}}\; , \nonumber \\
z^{\mathrm min} &=& \frac{x_T}{2}\,\left[ 
\frac{{\mathrm e}^{-\eta}}{x_2}+\frac{{\mathrm e}^{\eta}}{x_1}\right] \; .
\end{eqnarray}
 From these equations it follows that at central rapidities
$\eta\approx 0$ the momentum fractions $x_1$ and $x_2$ 
can become as small as roughly $p_T/\sqrt{s}$. In {\em forward} 
scattering, that is, at (large) positive $\eta$, the collisions 
become very asymmetric. In particular, $x_2$ may become fairly
small, whereas $x_1$ tends to be large. For forward kinematics
at {\sc Brahms} one has, typically, $p_T\sim 1.5$ GeV and $\eta =3.2$. 
This implies that $x_2$ may become as small as $\sim 3.5\times
10^{-4}$. However, in practice it turns out that such small $x_2$
hardly ever contribute to the cross section: if $x_2$ is so small, 
the hadron with transverse momentum $p_T$ can only be produced 
if both $x_1$ and $z$ are unity, where however the parton distributions
$f_a^{H_1} (x_1,\mu)$ and the fragmentation functions $D_c^h(z,\mu)$
vanish. This is an immediate consequence of kinematics, as 
demonstrated by Eqs.~(\ref{eq2}).  One can show that if the 
parton density $f_a^{H_1} (x_1,\mu)$ behaves at large $x_1$ 
as $(1-x_1)^{a_f}$ and $D_c^h(z,\mu)$ as $(1-z)^{a_D}$
(with some powers $a_f,a_D\gg 1$), the $x_2$-integrand
in Eq.~(\ref{eq1}) vanishes in the vicinity of 
$x_2^{\mathrm min}$ as $(x_2-x_2^{\mathrm min})^{a_f+a_D+1}$.
Therefore, contributions from very small $x_2$ are highly
suppressed. 

\begin{figure}[t]
\begin{center}
\epsfig{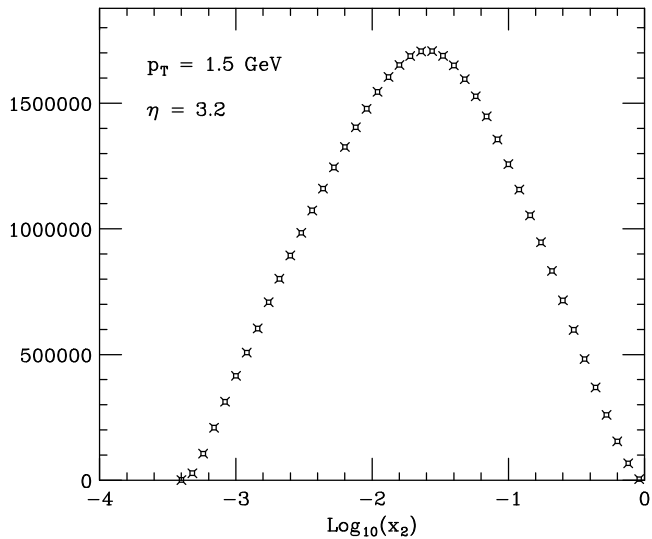}
\end{center}
\vspace*{-0.3cm}
\caption{\label{fig:fig1} \sf Distribution  in $\log_{10}(x_2)$
of the NLO invariant cross section $E d^3\sigma/dp^3$ at 
$\sqrt{s}=200$~GeV, $p_T=1.5$~GeV and $\eta=3.2$.}
\end{figure}

The question, then, remains of how small $x_2$ really is on average
for forward kinematics at RHIC. This is of course
relevant for judging various explanations for the suppression
of $R_{{\mathrm{dA}}}$ seen by {\sc Brahms}, in particular those relating to
saturation effects in the nucleus wave function~\cite{cgcmodels}. 
Figure~\ref{fig:fig1} shows the distribution of the cross section 
for $pp\to \pi^0X$ at $\sqrt{s}=200$~GeV, $p_T=1.5$~GeV, 
$\eta=3.2$, in bins of $\log_{10}(x_2)$. The overall 
normalization is unimportant of course; for definiteness 
we note that the sum of all entries shown in the plot yields 
the full NLO invariant cross section $E d^3\sigma/dp^3$ in pb/GeV$^2$. 
For the calculation we have chosen the CTEQ6M \cite{cteq6} 
parton distribution functions and the fragmentation functions 
of Ref.~\cite{kkp}. 
One can see that the distribution peaks
at $x_2>0.01$. There are several ways to estimate an 
average $\langle x_2 \rangle$ of the distribution. For example,
one may define $\langle x_2 \rangle$ in the standard way from 
evaluating the integral in Eq.~(\ref{eq1}) with an extra factor 
$x_2$ in the integrand, divided by the integral itself:
\begin{equation} \label{eq4av}
\langle x_2 \rangle \; \equiv  \frac{\int_{x_2^{\mathrm min}}^1 dx_2 \;
x_2 \; f_b^{H_2} (x_2,\mu) \ldots }{ \int_{x_2^{\mathrm min}}^1 dx_2 \;
f_b^{H_2} (x_2,\mu) \ldots}  \; 
 \; ,
\end{equation}
where the ellipses denote the remaining factors in Eq.~(\ref{eq1}).
Alternatively, one may simply determine $\langle x_2 \rangle$
as the median of the distribution, demanding that the
area under the distribution in Fig.~\ref{fig:fig1} to the
left of $\langle x_2 \rangle$ equals that to the right. 
Either way, one finds an average $\langle x_2 \rangle>0.01$, 
typically $0.03 - 0.05$ at this $p_T$ and $\eta$. 

\begin{figure}[t]
\begin{center}
\epsfig{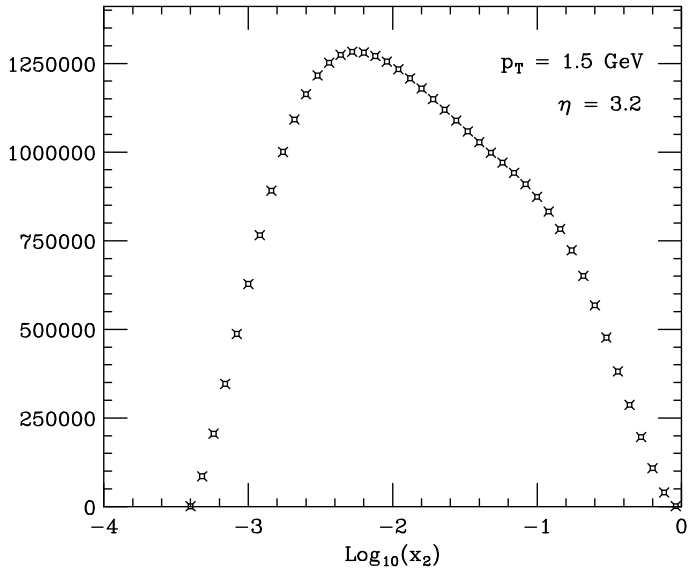}
\end{center}
\vspace*{-0.3cm}
\caption{\label{fig:fig2} \sf Same as Fig.~\ref{fig:fig1} but for
the GRV \cite{grv} parton distributions.}
\end{figure}

The precise shape of the distribution and the value of 
$\langle x_2 \rangle$ depend somewhat on the parton 
distributions (and, less so, on the fragmentation functions) 
chosen. We remind the reader that the distribution shown in 
Fig.~\ref{fig:fig1} is at $p_T=1.5$~GeV and that we have
chosen the factorization and renormalization scales
to be $\mu=p_T$. This means that we are using a 
fairly low scale in the parton densities.  At this
scale, the CTEQ6 densities, in particular the gluon, 
are still relatively flat towards small $x$. In order 
to estimate to what extent this influences
the distribution, we have calculated it for the 
GRV \cite{grv} parton distributions, which are
steeper at this scale. The corresponding histogram
in  $\log_{10}(x_2)$ is shown in Fig.~\ref{fig:fig2}. One can see
that as expected it peaks somewhat more to the left; 
nevertheless there is not much quantitative change in the
average $x_2$.  The full invariant cross section is about $15\%$
smaller than for the CTEQ6 set.
We have mentioned in the introduction that there are data from {\sc Star}
for the cross section for $pp\to \pi^0X$ in roughly this kinematic 
range \cite{star} which are in very good agreement with the
NLO calculation used in Fig.~\ref{fig:fig1}. 
This supports the view that the distributions
shown in Figs.~\ref{fig:fig1} and \ref{fig:fig2} are realistic.

\begin{figure}[t]
\begin{center}
\hspace*{8mm}
\epsfig{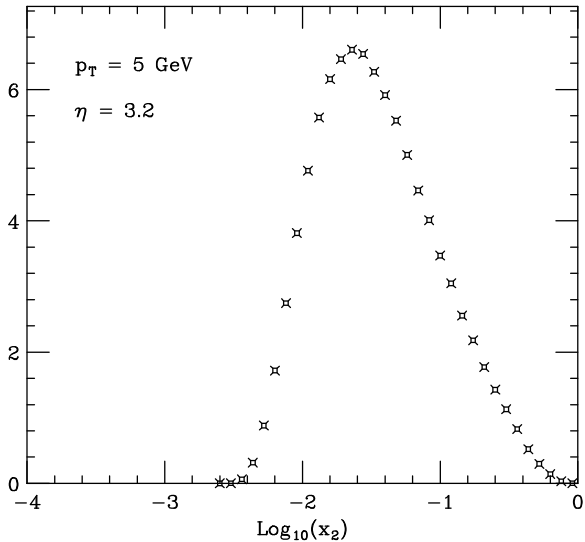}
\end{center}
\vspace*{-0.3cm}
\caption{\label{fig:fig3} \sf Same as Fig.~\ref{fig:fig1}, but for 
$p_T=5$~GeV.}
\end{figure}

Figure~\ref{fig:fig3} shows the $\log_{10}(x_2)$ distribution 
at $p_T=5$~GeV. At this $p_T$, one is closer to the
boundary of phase space given by the condition $x_T\cosh(\eta)= 1$,
where all momentum fractions $x_1, x_2, z$ are forced to 1.
The distribution in $x_2$ is therefore more ``squeezed''
and shifted to the right. The effect is countered to some
extent by evolution since at scale $\mu=5$~GeV the parton 
distributions are steeper than at $\mu=1.5$~GeV.

\begin{figure}[t]
\begin{center}
\vspace*{4mm}
\epsfig{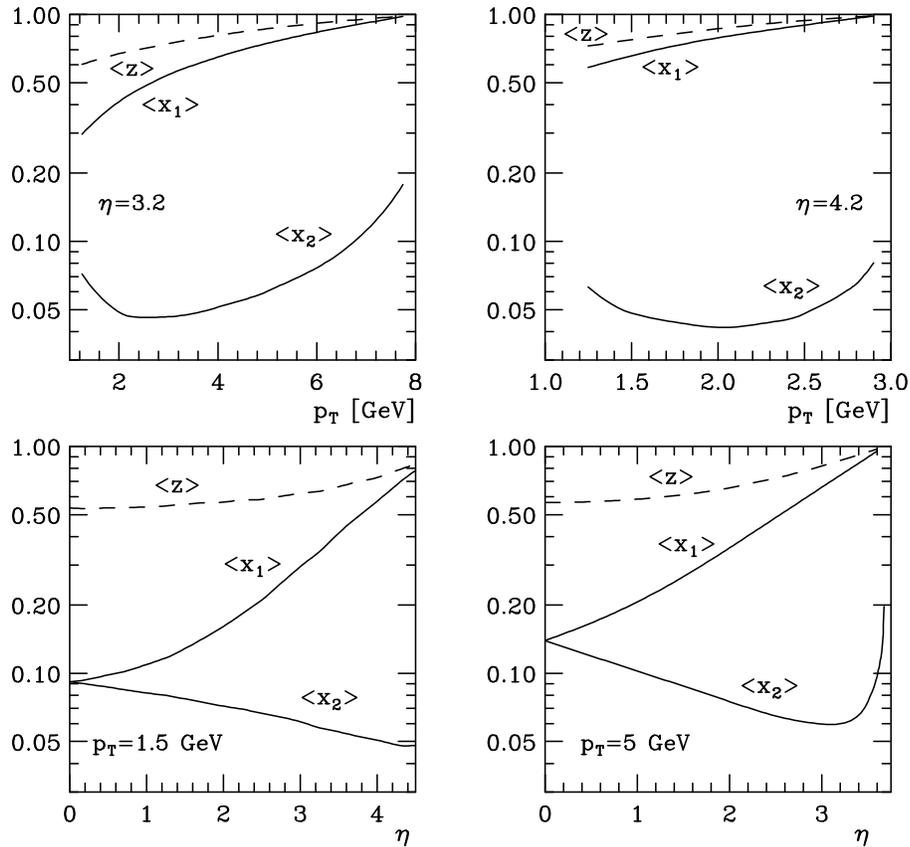}
\end{center}
\vspace*{-0.3cm}
\caption{\label{fig:fig4} \sf Averages of $x_1$, $x_2$, and $z$
in $pp\to \pi^0 X$ at RHIC, defined as in Eq.~(\ref{eq4av}), as 
functions of pion transverse momentum at forward rapidities (upper row), 
and of pion rapidity for fixed $p_T$ (lower row).}
\end{figure}

Finally, in Figure~\ref{fig:fig4} we present results for 
the averages of $x_1$, $x_2$, and $z$, as functions of 
pion transverse momentum and rapidity at $\sqrt{s}=200$~GeV. 
Here we have defined $\langle x_2 \rangle$ as in Eq.~(\ref{eq4av}),
with analogous definitions for $\langle x_1 \rangle $
and $\langle z \rangle $. 
The upper row shows results for fixed $p_T$ in forward
scattering. Besides $\eta=3.2$ as relevant for {\sc Brahms},
we have also extended the results to $\eta=4.2$ which
may be useful for future experimental studies. It becomes
evident that $\langle x_1\rangle $ and $\langle z\rangle $ 
are very large in forward scattering, as expected. $\langle z\rangle $ 
is particularly large because, on account of Eq.~(\ref{eq1}),
the single fragmentation function has to compete with two parton 
densities, each function being strongly suppressed at large
momentum fraction. As we have already seen in the 
histograms, Figs.~\ref{fig:fig1}--\ref{fig:fig3}, $\langle x_2\rangle$ 
is much smaller. As $p_T$ increases and the boundary of phase 
space is approached, all momentum fractions become larger and 
eventually converge to unity. We also note an unexpected upturn
of $\langle x_2\rangle$ toward smaller $p_T$. We have not been 
able to identify this effect as resulting from any straightforward 
origin. The precise small-$x$ behavior of the parton distributions 
at the rather low scales involved here 
plays a role (however, the effect also occurs for the steeper GRV 
distributions). The structure of the cross section formula in 
Eq.~(\ref{eq1}) itself is also partly responsible. In the lower part of
Fig.~\ref{fig:fig4} we show the averages as functions of rapidity for 
two fixed values of $p_T$. At $\eta=0$ one obviously starts 
from $\langle x_1\rangle=\langle x_2\rangle$; with increasing $\eta$
the two momentum fractions become very different.
Toward $\eta=\cosh^{-1}(1/x_T)$ they again both tend to unity; for 
$ \langle x_2\rangle$ this happens rather late.

\section{Influence of leading-twist nuclear shadowing}
\label{nuclshad}

In dA collisions, nuclear effects will alter the distribution 
in $x_2$ as well as the full cross section. Shadowing effects at 
small-$x_2$, $x_2 < 0.05-0.1$,  will lower the cross section and will lead to
yet higher average $x_2$ being probed. There will also be enhancements
at larger $x_2$, $0.05-0.1 < x_2 < 0.2$, associated with yet another
coherent nuclear effect, antishadowing.  This will be followed
by the suppression related to the EMC effect for $0.2 < x_2 < 0.8$, and by
the subsequent enhancement explained by the Fermi motion for $x_2 > 0.8$.

Since in the {\sc Brahms} kinematics the average $x_2$ is near 0.01, 
the principal nuclear effect is shadowing.
We investigate its role in the interpretation of the {\sc Brahms} data by
considering leading-twist shadowing, using the parameterization of nuclear
parton distribution functions (nPDFs) derived in~\cite{sh1,sh2,sh3}.
Unlike most other sets of nPDFs~\cite{ds,other}, these functions 
have a rather rapid and strong onset of shadowing toward small-$x$, 
so they may serve as a good tool for studying the ``maximally
possible'' effects of leading-twist nuclear shadowing
in forward dA scattering. The recent study~\cite{rvogt}
has also employed the nPDFs of~\cite{sh1,sh2,sh3}, albeit 
only in the framework of a lowest order (LO) calculation.

We note that we will neglect any nuclear effects in the deuteron, 
for which we just use $d=(p+n)/2$. 
As follows from our analysis of the average $x_1$,
the deuteron parton distributions are sampled at values of  
$x_1$ in the domain of the EMC effect. Therefore, the approximation
$d=(p+n)/2$ is valid to a few per cent accuracy, as can be 
estimated as follows.  
The CTEQ fits use the neutron structure function extracted
from the deuteron data within the framework of the convolution 
approximation (Fermi motion). The convolution model gives for the 
structure function ratio $R=2F_2^d/(F_2^p+F_2^n)$
the values of 0.99 for $x\sim 0.5$ and of $1.03-1.05$ for $x\sim 0.7$.
As a result, our $d=(p+n)/2 $ approximation overestimates the true 
deuteron parton distributions by about $\sim 1\%$ at $x\sim 0.5$
and underestimates them by a few percent at $x\sim 0.7$.
Since large $x_1$ are important in our calculations, proper 
account of this would make the effect we will discuss in the
next section even slightly bigger. Note that for heavier nuclei 
the convolution model contradicts the EMC effect. However here 
we are using it to ``restore'' the deuteron structure function 
within the  procedure used to extract the neutron 
structure function; see \cite{FS} for an extensive discussion
of nuclear effects in the deuteron parton densities.

Let us now briefly describe the approach of~\cite{sh1,sh2,sh3}
for deriving nPDFs. Leading-twist nuclear shadowing is
obtained using Gribov's theorem~\cite{gribov} 
relating nuclear shadowing to diffraction, Collins's QCD factorization 
theorem for hard diffraction in DIS~\cite{collins}, and the QCD 
analysis of hard diffraction measured at HERA in terms of diffractive
parton distribution functions of the proton. Operationally, the nPDFs 
are first derived at the initial scale $Q_0$=2 GeV and for $10^{-5} 
\leq x \leq 1$. Standard (NLO) DGLAP evolution is then used to obtain the 
nPDFs for $Q^2 > Q_0^2$. 

Analyses of DIS by both H1~\cite{h1} and ZEUS~\cite{zeus} demonstrate that 
diffraction constitutes $\approx 10$\% of the total cross section in 
the quark-dominated channel and as much as $\approx 30$\% in the 
gluon channel. As a result, it is found in~\cite{sh1,sh2,sh3} 
that the effect of nuclear shadowing in nPDFs is large and, even 
more strikingly, much stronger in the gluon nPDF at small $x$ 
than in the quark nPDFs.

Conservation of the baryon number and the 
momentum sum rule then require that the depletion of nPDFs at small 
values of $x$, $x < 0.01$, be accompanied by a certain enhancement at 
$0.05 < x < 0.2$. The transition from shadowing to enhancement, 
and the enhancement itself, are not described by the Gribov theorem 
and, hence, can be only modeled. Using the available fixed-target 
nuclear DIS data~\cite{disn} as a guide, the ``standard'' scenario of 
\cite{sh1,sh2,sh3} assumes that the transition from nuclear shadowing 
to the enhancement takes place at $x=0.1$ for quark nPDFs, and 
at $x=0.03$ for the gluon nPDF. In the following, we will refer
to this set of nPDFs as ``shadowing 1''.
\begin{figure}[t]
\begin{center}
\epsfig{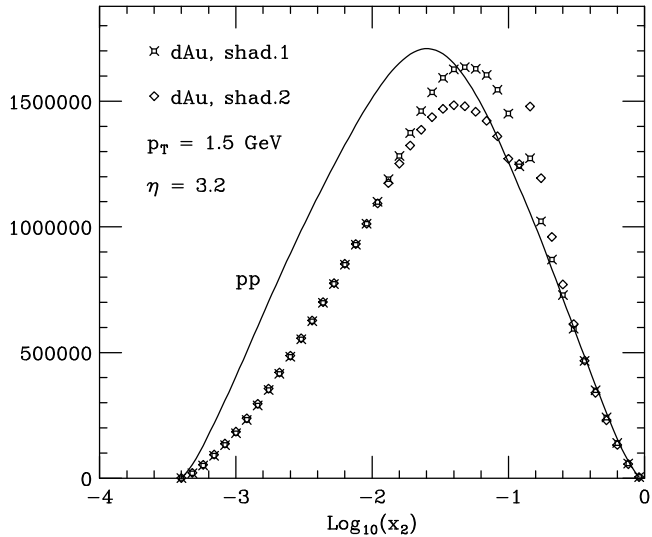}
\end{center}
\vspace*{-0.3cm}
\caption{\label{fig:fig5} \sf Same as Fig.~\ref{fig:fig1}, but
also showing the result for ${\mathrm{dAu}}$ scattering using the shadowing
of Refs.~\cite{sh1,sh2,sh3}, and a more extreme ansatz for shadowing
(see text). The ``spikes'' in the distributions at $\log_{10}(x_2)
\approx -0.8$ are artifacts of the implementation of antishadowing
in the nPDFs of~\cite{sh1,sh2,sh3}. For better comparison we 
have displayed the result of Fig.~\ref{fig:fig1} by a solid line.}
\end{figure}

Figure~\ref{fig:fig5} shows the distribution of the NLO cross section
for ${\mathrm{dAu}}\to \pi^0X$ in $\log_{10}(x_2)$, using shadowing 1.
For comparison, we also display the previous result for $pp\to \pi^0X$ 
of Fig.~\ref{fig:fig1} (solid line). A clear
shift in the distribution to larger $x_2$ is visible. At small
$x_2$, there are significant shadowing effects, while at
large $x_2$ there is a small contribution from antishadowing. 
It is evident from comparison of the areas underneath the distributions
that the net effect on the ${\mathrm{dAu}}$ cross section will be a 
decrease. However, one can anticipate that the decrease will
be rather moderate: while nuclear shadowing does deplete
the ${\mathrm{dA}}$ cross section compared to the $pp$ cross section, 
the probed values of $x_2$ are clearly not small enough to deliver
a significant nuclear shadowing effect. 

The kinematics for forward scattering at {\sc Brahms} 
mostly corresponds to values of $x_2$ in the transition
region between shadowing and antishadowing, where the predictions for
nPDFs are rather uncertain. Therefore, in addition to the standard 
scenario (``shadowing 1'') of nuclear shadowing, we have also explored 
an option for which nuclear shadowing in the gluon channel is increased 
by extending it up to $x=0.1$, similarly to the shadowing in the quark 
densities. We refer to the resulting set of nPDFs as ``shadowing 2''.
The corresponding $\log_{10}(x_2)$ distribution of the 
cross section for ${\mathrm{dAu}}\to \pi^0X$ is also displayed
in fig.~\ref{fig:fig5}. Compared to shadowing 1, there is only 
a small modification of the distribution, which will lead 
to a very slight further
suppression of the dA cross section. It is worth emphasizing 
that one can hardly increase the amount of gluon shadowing at 
$x \sim 10^{-3}$ since here there are constraints from $J/\psi$ 
data~\cite{jpsi}. The model of~\cite{sh1,sh2,sh3} gives a
reasonable description of the observed suppression by a factor 
$ \sim 2$, which would be spoilt by a much stronger gluon shadowing.
At the same time, as soon as the amount of gluon shadowing at 
$x\sim 10^{-3}$ is fixed, the gradual decrease of shadowing 
with increasing $x$ is automatic as a consequence of the decrease 
of the coherence length $\propto 1/m_N x$ and of a smaller probability 
for diffraction.
Therefore, we conclude that the standard effect of leading twist 
nuclear shadowing will at best be able to explain only a small 
fraction of the dramatic suppression of the spectra of charged 
hadrons at forward rapidities observed by {\sc Brahms}. This 
statement is in line with the LO result in the revised version 
of~\cite{rvogt}.

Figure~\ref{fig:fig5} shows the corresponding results for $p_T=5$~GeV.
Here, larger $x_2$ are probed, and only slight antishadowing effects 
appear. 
\begin{figure}[t]
\begin{center}
\epsfig{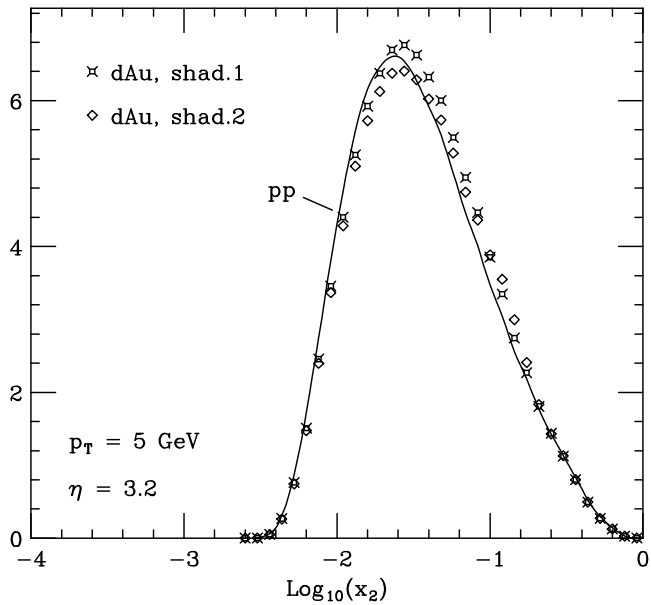}
\end{center}
\vspace*{-0.3cm}
\caption{\label{fig:fig6} \sf Same as Fig.~\ref{fig:fig5}, but
at $p_T=5$~GeV.}
\end{figure}

So far we have only considered $\pi^0$ production as an example. 
This is however not really appropriate for a comparison with the 
{\sc Brahms} data which, at forward rapidities, are for negatively 
charged hadrons $h^-$. As we will now show, for the case of $h^-$, 
even in presence of the shadowing effects just discussed, the 
leading-twist NLO calculation predicts an {\it enhancement}, rather 
than a suppression, of $R_{{\mathrm{dA}}}$.

\section{Isospin considerations for the ratio of ${\mathrm{dA}}$ 
and $pp$ cross sections}
\label{isospin}
We now consider the ratio $R_{{\mathrm{dA}}}$ of single-inclusive 
hadron cross sections in ${\mathrm{dA}}$ and $pp$ scattering. 
The {\sc Brahms} experiment has presented data~\cite{brahms}
for $R_{{\mathrm{dA}}}$ as a function of hadron transverse
momentum $p_T$, in four different bins of hadron pseudorapidity
$\eta$, with central values $\eta=0,1,2.2,3.2$. 
{\sc Brahms} sees a significant suppression of the ratio with 
increasing $\eta$. 

\begin{figure}[t]
\begin{center}
\epsfig{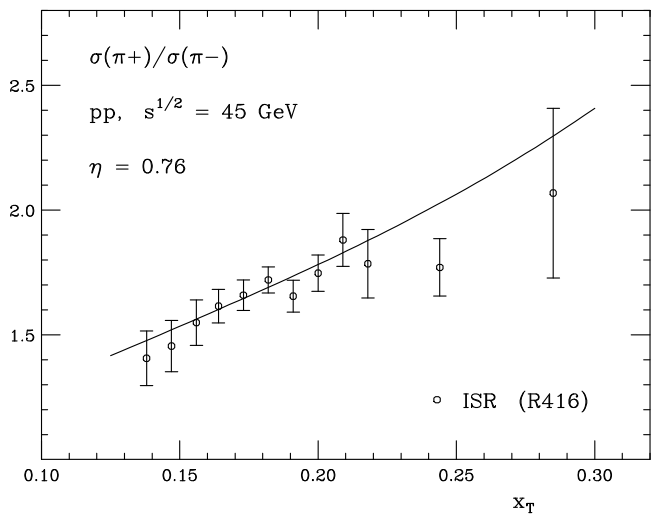}
\end{center}
\vspace*{-0.3cm}
\caption{\label{fig:fig7} \sf Ratio of $pp\to \pi^+ X$ and 
$pp\to \pi^- X$ cross
sections at $\sqrt{s}=45$~GeV and scattering angle
$50^o$ (corresponding to pseudorapidity $\eta=0.76$), 
as a function of $x_T\equiv 2 p_T/\sqrt{s}$. The
data points are from measurements at the ISR \cite{isr}. The
curve shows the result of the NLO calculation, using
the fragmentation functions of~\cite{kretzer}.}
\end{figure}

While {\sc Brahms} measures inclusive charged hadrons,
$\left( h^+ + h^- \right)/2$, at central rapidities 
($\eta=0$ and 1), their $R_{{\mathrm{dA}}}$ data at forward
rapidities refer only to {\it negatively} charged hadrons
$h^-$. This has profound consequences. To see this, let
us assume for the moment that pions dominate the 
spectrum of observed high-$p_T$ hadrons. Negatively 
charged pions are produced more rarely than positively 
charged ones in $pp$ collisions, due to the up-quark 
dominance in the proton. An example for this is shown
in Fig.~\ref{fig:fig7}, where we display data for the $\pi^+/\pi^-$ ratio
from the ISR~\cite{isr} at $\sqrt{s}=45$~GeV. We also show
the result of the NLO calculation, using the fragmentation
functions of Ref.~\cite{kretzer}, which provides separate sets 
for negatively and positively charged pions. [We note that
there are also $\pi^+/\pi^-$ data at $\sqrt{s}=62$~GeV~\cite{isr1}
which lie lower and are in less impressive agreement with NLO pQCD.] 
As Fig.~\ref{fig:fig7} shows, there is clear excess of positive
pions over negative ones. In contrast, isospin considerations
imply that $\pi^+$ and $\pi^-$ are produced practically equally 
in dAu collisions. Therefore, one expects $R_{{\mathrm{dA}}}$ for
negatively charged pions to be intrinsically enhanced, if it 
is normalized by the $pp$ cross section and not, for example,
by the $dp$ one.

\begin{figure}[t]
\begin{center}
\epsfig{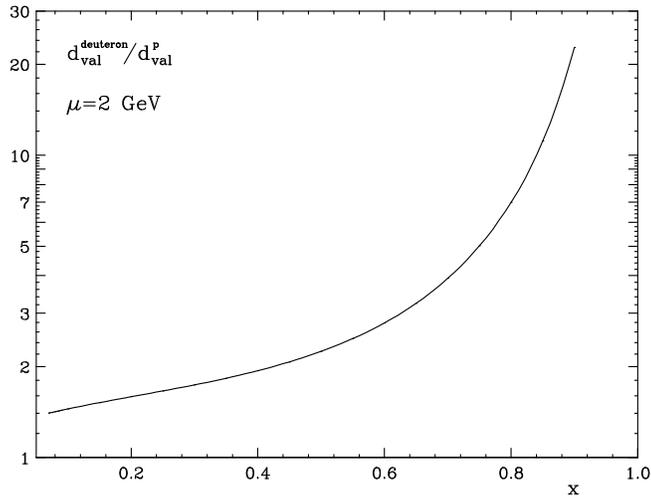}
\end{center}
\vspace*{-0.3cm}
\caption{\label{fig:fig8} \sf Ratio $d_{val}^{deuteron}(x)/d_{val}^p(x)$
as a function of $x$ at the scale $\mu=2$~GeV, as given by the 
CTEQ6M set. The deuteron's $d$-valence distribution has been
estimated according to Eq.~(\ref{dval}).}
\end{figure}

\begin{figure}[t]
\begin{center}
\epsfig{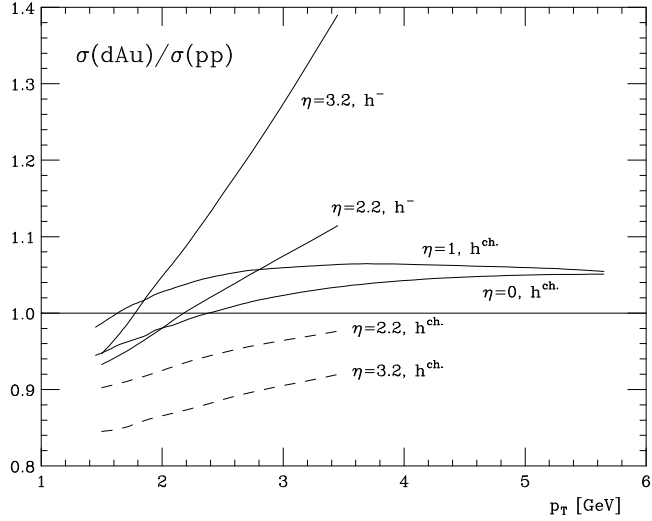}
\end{center}
\vspace*{-0.3cm}
\caption{\label{fig:fig9} \sf Ratio $R_{{\mathrm{dA}}}$
of cross sections for ${\mathrm{dAu}}\to h X$ and
$pp\to h X$ as a function of transverse momentum at various 
rapidities relevant to the {\sc Brahms} experiment. As 
in experiment, we have considered production of {\it summed}
charged hadrons, $h^{ch.}\equiv(h^+ + h^-)/2$ for $\eta=0,1$ and
{\it negatively} charged hadrons $h^-$ for $\eta=2.2$ and $3.2$.
For comparison, the dashed lines show the result for summed
charged hadrons at $\eta=2.2$ and $3.2$. We have used the 
``shadowing 1'' nPDFs for the gold nucleus. The 
fragmentation functions are from~\cite{kretzer}; 
we have found that for the case of summed charged 
hadrons using the set of \cite{kkp} does not alter our 
results by more than a few per cent.}
\end{figure}

To go into a little more quantitative detail, we recall that at 
forward rapidities the partonic collisions are very asymmetric. 
Large contributions to the scattering come from situations in which 
a large-$x_1$ valence quark in the ``projectile'' (i.e., in the deuteron, 
or in one of the protons) hits a small-$x_2$ gluon in the ``target'' (i.e., 
in the gold nucleus or in the other proton). The underlying (LO) subprocess 
is then the quark-gluon Compton process $qg\to qg$. For negatively
charged pions one then expects that down quarks play a 
particularly important role in the Compton process, since 
both the ``projectile'' and the produced $\pi^-$ have a $d$ 
valence quark. To a good approximation (see the previous section), 
the deuteron's $d$ valence density is given by
\begin{equation} \label{dval}
d_{val}^{deuteron}\; =\; \frac{1}{2} \, \left( 
d_{val}^p \, + \, d_{val}^n \right) \;=\;
\frac{1}{2} \, \left( 
d_{val}^p \, + \, u_{val}^p \right) \; ,
\end{equation}
where we have used isospin invariance to relate the valence-$d$
distribution in the neutron to the valence-$u$ in the proton. 
Due to the up-quark excess in the proton, the distribution 
in Eq.~(\ref{dval}) becomes much larger than the proton's $d$-valence 
distribution at high $x$, as shown in Fig.~\ref{fig:fig8},
resulting in an enhancement in $R_{{\mathrm{dA}}}$. 
Of course, other scattering channels will contribute as well 
and dilute this valence effect. In addition, {\sc Brahms}
does not measure only pions, but inclusive charged hadrons,
$h^{\pm} = \pi^{\pm} + K^{\pm} + \stackrel{(-)}{p}+\ldots$.
Nevertheless, when changing to charged-hadron fragmentation
functions as given by~\cite{kretzer}, we find that the 
difference in deuteron and proton valence densities continues 
to play an important role in the forward production 
of negatively charged hadrons. This is demonstrated 
by Fig.~\ref{fig:fig9}. The solid lines show the ratio 
$R_{{\mathrm{dA}}}$ at $\eta=0$ and $1$ for {\it summed}
charged hadrons $\left( h^+ + h^- \right)/2$, and 
at $\eta=2.2$ and $3.2$ for {\it negatively charged} 
hadrons, exactly corresponding to the {\sc Brahms} conditions.
We have used the ``shadowing 1'' set described in the previous section. 
As shown in sec.~\ref{elementary}, when going from $\eta=0$ to 
$\eta=1$, the average $x_2$ probed slightly
decreases, and shadowing effects start to become visible 
at the smaller $p_T$. At $\eta=2.2$ and $\eta=3.2$ the $x_2$
become yet smaller, but since now negatively charged hadrons
are measured,  the valence effect discussed above
outweighs any stronger shadowing, and in fact the 
ratio $R_{{\mathrm{dA}}}$ strongly increases with $p_T$
because larger and larger $x_1$ become relevant. For
comparison we also show in Fig.~\ref{fig:fig9} the results
at $\eta=2.2$ and $\eta=3.2$ for {\it summed} charged hadrons
(dashed lines). For these, the effective valence densities
in the proton and deuteron are the same, and the enhancement
seen for negative hadrons disappears. Shadowing effects of 
up to $15\%$ are visible 
then, as expected from Fig.~\ref{fig:fig5}, and as also 
found in Ref.~\cite{rvogt} where only summed charged hadrons were
considered. It is remarkable that for the highest $p_T\sim 3$~GeV at
$\eta=3.2$ the curve for $h^-$ is enhanced by about a factor
1.5 with respect to the one for summed charges.

Our analysis so far has been entirely based on a NLO 
pQCD leading-twist calculation and the use of fragmentation 
functions extracted from $e^+e^-$ annihilation data. We expect that 
even if nonperturbative phenomena are important in the 
kinematic regime explored by {\sc Brahms}, the enhancement
in $R_{{\mathrm{dA}}}$ resulting due the different ``projectiles''
in the numerator and denominator will persist. Here, our 
reasoning is as follows: ISR data (see~\cite{Singh:1978ra}
and references therein) on $\pi^{\pm}$ production at $p_T\sim 1$~GeV 
and large Feynman-$x_F\sim 0.4$ were found to be consistent with 
$d\sigma^{pp\to\pi^+}/dx_F\propto u(x_F)$, 
$d\sigma^{pp\to\pi^-}/dx_F\propto d(x_F)$, where $u$ and $d$ are
typical densities for up and down valence quarks. Such spectra
in $x_F$ are much harder than the ones pQCD would give, indicating
the presence of a nonperturbative production mechanism. Nonetheless,
the ratio of $\pi^+$ to $\pi^-$ cross sections is of a similar magnitude 
as the one in the perturbative regime shown in Fig.~\ref{fig:fig7}.
Adding the fact that the yields of $\pi^+,\pi^-, \pi^0$ in
dA scattering should be equal because of isospin, irrespective
of the production mechanism, we are led to the conclusion 
that $R_{{\mathrm{dA}}}$ for negatively charged hadrons should 
be enhanced even if nonperturbative effects dominate.
Note here that nonperturbative effects in the fragmentation region 
are known to be consistent with the   Feynman scaling. Hence if
nonperturbative effects are important they should lead to about the same $\pi^+/\pi^-$
ratio in $pp$ scattering  for the same $x_F$ at ISR and at RHIC.

We finally note that a potential caveat to our main finding
in Fig.~\ref{fig:fig9}
comes from a further set of {\sc Brahms} data. While we have 
mentioned that {\sc Brahms} measures all hadrons and not
just pions, we have generally assumed that pions dominate
the observed hadron spectrum. However, preliminary data from
{\sc Brahms}~\cite{brahms1} show that the cross section for ${\mathrm{d Au}}
\to h^+X$ becomes significantly larger than that for
${\mathrm{d Au}}\to h^- X$ at $p_T>1$~GeV. At $p_T=3$~GeV,
they observe about three times as many $h^+$ as $h^-$.
As we have pointed out before, isospin excludes that such 
an excess could be due to pions: 
$\sigma({\mathrm{d Au}} \to \pi^+X)=\sigma({\mathrm{d Au}} 
\to \pi^-X)$, up to corrections of a few per cent
related to the fact that the gold nucleus is not isoscalar.
Standard sets of fragmentation functions do not predict large
contributions from kaons and protons, and in the NLO calculation one
ends up with $\sigma({\mathrm{d Au}} \to h^+X)$ at most only
$10\%$ larger than  $\sigma({\mathrm{d Au}} \to h^-X)$.
It is hard to conceive that proton production could lead to a
large enhancement of $h^+$ over $h^-$ (that even increases with $p_T$)
but, barring any experimental systematic problem, this
appears to be the conclusion at present. 
If the final {\sc Brahms} data continue to show this
large excess, it will be a challenge to understand it in 
terms of a nonperturbative effect. Such an effect could, perhaps,
result from coalescence of quarks from the incoming nucleon 
with other partons, to form a baryon. For this to work, the
quarks would need to experience large transverse ``kicks'' 
and would need to lose a significant fraction of their
momentum. Such a possibility could be connected to the expectations
of a very strong suppression of the forward nucleon spectrum in  
central nucleon-nucleus collisions \cite{berera,dumitru}.

\section{Conclusions and outlook}
\label{conclusions}

We have shown that there is an intrinsic enhancement in the 
ratio $R_{{\mathrm{dAu}}}$ for negatively charged hadrons, 
simply because of the different nature of the ``projectile'' 
(deuteron vs. proton) in the numerator and denominator of 
$R_{{\mathrm{dAu}}}$. 
In the light of Fig.~\ref{fig:fig9} the significant suppression 
seen by {\sc Brahms} at $\eta=2.2$ and $\eta=3.2$ is 
even more striking than usually appreciated. The effect we have found
has not been taken into account in any previous theoretical
study~\cite{cgcmodels,qv,rvogt} of the data, to our knowledge. 
We expect that future data for $R_{{\mathrm{dAu}}}$ for {\it summed}
charged hadrons in this kinematic regime will show
an even stronger suppression than observed for $h^-$,
roughly by a factor 1.5. The same should happen if 
$R_{{\mathrm{pAu}}}$, rather than  $R_{{\mathrm{dAu}}}$, were
measured for $h^-$.

Because of the effect, it is entirely impossible to explain 
the suppression in $R_{{\mathrm{dAu}}}$ at forward rapidities
by a conventional modification of the leading-twist parton 
densities in nuclei. However, {\it even if} we disregard the
effect, nuclear leading-twist shadowing plays a rather
unimportant role, giving at most a suppression of $15\%$. 
The reason for this is that parton momentum fractions in
the gold nucleus are not very small on average even for
forward kinematics, as we have shown. In other words, a large nuclear 
contribution originates from a range of $x$ where nuclear effects 
are known to be small (or even antishadowed). This generally sets
severe limitations on the ability of {\em any} initial-state small-$x$ 
effects to explain the observed suppression.

We have also mentioned that the fact that {\sc Brahms} observes
the cross section for ${\mathrm{d Au}} \to h^+X$ to be significantly 
larger than that for ${\mathrm{d Au}}\to h^- X$ at $p_T>1$~GeV,
indicates the presence of sizable nonperturbative contributions
possibly related to protons. It appears likely that
mechanisms responsible for an enhancement of proton production
then also play a role in the observed suppression of $R_{{\mathrm{dAu}}}$.
Nonperturbative production of pions, too, could play a role:
coalescence effects involving spectator partons are likely to be 
strongly suppressed in p(d)A collisions as compared to the $pp$ 
case~\cite{berera}. This suppression is further enhanced when 
energies are large enough to resolve the small-$x$ high gluon 
densities~\cite{dumitru}.
Such non-leading twist effects should decrease with increase of 
the transverse momentum of the pion, which is consistent with the trend in 
the data on $R_{{\mathrm{dAu}}}$. On the other hand, the observed excess of 
${\mathrm{d Au}} \to h^+X$ over ${\mathrm{d Au}} \to h^-X$
actually {\it in}creases  with $p_T$, which is quite challenging
to understand. Note also that the parton energy losses that would 
be necessary to reproduce the Brahms effect appear to be rather 
large: about 10\% energy loss would be needed if it occurred only
in the initial state before the interaction. If one assumes 
that the rate of energy loss is the same in the initial and final 
states, a loss of about 3\% is necessary. For the kinematics
relevant here, the suppression is more sensitive to losses in the 
final state since the average $\langle z \rangle$ for fragmentation 
was found to be substantially closer to one than the averages 
$\langle x_{1,2} \rangle$ in the parton densities 
(see Fig.~\ref{fig:fig4}).

\begin{figure}[h]
\begin{center}
\epsfig{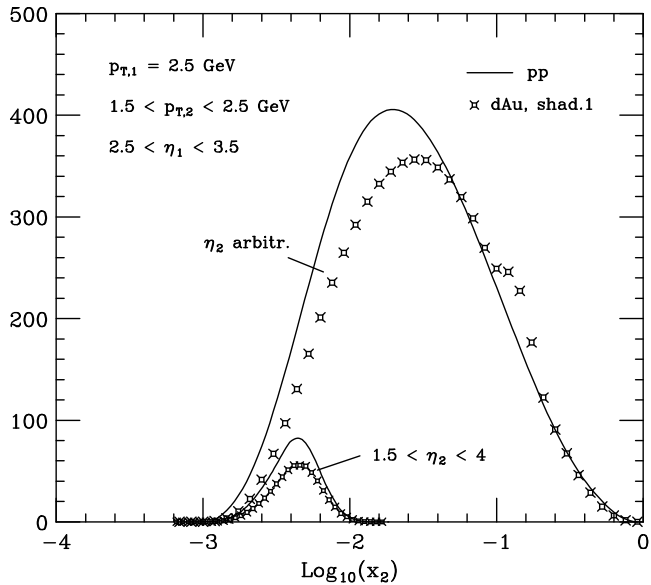}
\end{center}
\vspace*{-0.3cm}
\caption{\label{fig:fig10} \sf LO distributions in $\log_{10}(x_2)$
of the cross section $d\sigma/dp_{T,1}$ for $pp\to \pi^0 \pi^0 X$
and ${\mathrm{dAu}}\to \pi^0 \pi^0 X$
production at $\sqrt{s}=200$~GeV. The kinematic variables
have been chosen as described in the text. Solid lines are
for $pp$ collisions and histograms are for dAu, using ``shadowing 1''.
The higher-lying histograms are for the case of ``arbitrary''  
(i.e., unconstrained) $\eta_2$, the lower ones are for 
$1.5\leq\eta_2\leq 4$.}
\end{figure}

To further investigate experimentally the origin of the suppression in
$R_{{\mathrm{dAu}}}$ it would be useful to perform measurements of 
dihadron production. Within LO kinematics, the pseudorapidities
of the two hadrons are related by $\eta_1+\eta_2=\ln(x_1/x_2)$.
One may therefore single out contributions from small $x_2$ by
demanding that both hadrons be rather forward. Fig.~\ref{fig:fig10}
shows this for a sample calculation. We assume that a ``trigger''
hadron ($\pi^0$) is detected with transverse momentum $p_{T,1}=2.5$ and
forward rapidity $2.5\leq\eta_1\leq 3.5$. Let the second $\pi^0$
have $1.5\; \mathrm{GeV}\,\leq p_{T,2}\leq p_{T,1}$. Without 
any restriction on the rapidity $\eta_2$ of the second hadron,
one then obtains the higher $\log_{10}(x_2)$-distributions 
in Fig.~\ref{fig:fig10}. As expected, these look very much like 
the single-hadron distributions shown in Sections~\ref{elementary} 
and~\ref{nuclshad}. If now the second hadron is also in the 
forward region at $1.5\leq\eta_2\leq 4$, the lower 
distribution is obtained, which is entirely located at
$x_2\leq 0.01$. We also show in Fig.~\ref{fig:fig10}
the corresponding results for dAu collisions, using 
our ``shadowing 1''. The shadowing effects are much more
relevant for the double-forward distribution, as expected.  
We note that the distributions in Fig.~\ref{fig:fig10}
are normalized such that they sum to the cross section
$d\sigma/dp_{T,1}$ in pb/GeV. It is evident that there is a 
significant decrease in rate for two forward hadrons.
The results shown in Fig.~\ref{fig:fig10} are only LO. The NLO 
corrections are expected to be sizable; they are available~\cite{owens}.
The two hadrons we have studied in Fig.~\ref{fig:fig10} would
be nearly back-to-back in azimuthal angle. Further insights into
the dynamics may be gained by studying back-to-back azimuthal
correlations~\cite{khlev,star1}.  We finally note that
small-$x_2$ effects might also become more readily accessible
in conceivable future $pA$ collisions after increase of the
proton energy to 250~GeV at a later stage of RHIC operations.
%

\section*{Acknowledgments}
%
We are grateful to D.\ de Florian, B.\ J\"{a}ger, and 
M.\ Stratmann for valuable help, and to L.~Frankfurt, 
D.~Kharzeev, S.\ Kretzer, K.\ Tuchin, and R.\ Venugopalan 
for useful discussions and comments. V.G. is supported by Sofia 
Kovalevskaya Program of the Alexander von Humboldt Foundation.
M.S.'s research was supported by the DOE grant  No. DE-FG02-93ER40771.
W.V.\ is grateful to RIKEN, Brookhaven National Laboratory and the U.S.\
Department of Energy (contract number DE-AC02-98CH10886) for
providing the facilities essential for the completion of this work.

%

\end{document}